\documentclass[12pt,a4paper]{article}

\begin{document}

\title{Earth and Moon orbital anomalies\\
{\small \it Si non \`{e} vero, \`{e} ben trovato }}
\author{Ll. Bel\thanks{e-mail:  wtpbedil@lg.ehu.es}}

\maketitle

\date{}

\begin{abstract}

A time-dependent gravitational constant or mass would correctly describe the suspected increasing of both: the Astronomical unit and the eccentricity of the Lunar orbit around the Earth

\end{abstract}

{\it 1.- The model}

\vspace{1cm}
The gravitational model below, although it was initially motivated by a desire to show the potentiality of time dependent solutions of Einstein's equations, is essentially a pure Newtonian-like one, and differs from the classical model by the very simple substitution:

\begin{equation}
\label{G}
\mu\rightarrow (1-p\,t)\mu, \ \ \ \mu=Gm
\end{equation}
where\,\footnote{$p$ in this paper corresponds to the product $3p$ of \cite{Bel}. Similar substitutions have been considered before: they are reminded in the introduction of \cite{Duval}.} $p$ is a constant whose tentative value I assume to be $1.5\times 10^{-20} s^{-1}$, and where $p\,t$ is supposed to be small enough so that $(p\,t)^2$  is negligible  during the whole duration of the processes to be considered below.

Using obvious assumptions and notation, the equations of motion of a point body of unit mass moving in the gravitational field created by a point source of mass $m$  may thus be derived from the time dependent Hamiltonian:

\begin{equation}
\label{H}
H=\frac12(\dot r^2+r^2\dot\varphi^2)-(1-p\,t)\frac{\mu}{r}, \ \ \ \mu=Gm
\end{equation}
that leads to the following equations:

\begin{equation}
\label{Equ1}
\ddot{r}-r\dot{\varphi}^2=-\frac{\mu}{r^2}(1-p\,t)
\end{equation}

\begin{equation}
\label{Equ2}
\dot J=2\dot{r}\dot{\varphi}+r\ddot{\varphi}=0 \ \ \hbox{with} \ \ J=r^2\dot\varphi
\end{equation}
as well as:

\begin{equation}
\label{dH}
\dot H=\frac{p\mu}{r}
\end{equation}

{\it 2.- Increasing of the astronomical unit and the Moon to Earth distance}

\vspace{1cm}
The following example assumes that both the Earth and the Sun can be dealt with as two point bodies and that the Earth deviates from a circle of radius $r=a$, the astronomical unit, at time $t$ by a small constant amount $\delta r$ after a short interval of time $\delta t$. More precisely, we assume that at the present epoch we have:

\begin{equation}
\label{r1}
r=a     \quad  \ddot r=0
\end{equation}
Using the second assumption above and the definition of $J$ we get:

\begin{equation}
\label{r2}
r=\frac{J^2}{\mu}(1+p\,t) \ \ \hbox{and so:} \ \ \dot r=p\frac{J^2}{\mu}=p\,a
\end{equation}
that leads to the following result:

\begin{equation}
\label{dota}
\dot a=p\,a =0.07\, m\,yr^{-1}
\end{equation}

{\it Mutating mutandis} the Sun by the Earth , the Earth by the Moon and the astronomical unit {\bf a} by the mean distance b of the Moon from the Earth I get:

\begin{equation}
\label{bu}
\dot b=p\,b =0.000182\, m\,yr^{-1}
\end{equation}

{\it 3.-Increasing of the Earth around the Sun and  the Moon around the Earth orbits eccentricities}

\vspace{1cm}

Using the classical formula satisfied by the eccentricity $e$ :

\begin{equation}
\label{ecc}
e=\sqrt{1+\frac{2HL^2}{\mu^2}}
\end{equation}
where the kinetic moment $J$ is constant and the energy $H$ given by (\ref{H}) is time dependent, we obtain:

\begin{equation}
\label{dotecc}
\dot e=\frac{J^2}{e\mu^2}\dot H  \ \ \hbox{ or using (\ref{dH})} \ \ \dot e=\frac{J^2}{e\mu}\frac{p}{a}
\end{equation}
With $m$ being the mass of the Sun, {\bf a} the astronomical unit and $e$ the present eccentricity of the orbit of the Earth,  the result is:

\begin{equation}
\label{dteE}
\dot e=2.81\times 10^{-11}\,yr^{-1}
\end{equation}

{\it  Mutating mutandis} the corresponding result giving the increasing of the eccentricity of the Lunar orbit is:

\begin{equation}
\label{dteM}
\dot e=8.73\times 10^{-12}\,yr^{-1}
\end{equation}

{\it Conclusion}

\vspace{1cm}
From the four results here mentioned, $\dot{\bf a}$, {\bf e}, ${\dot b}$, and $\dot{\bf e}$, only the first and the fourth have been well documented, and surprisingly both can be derived using the same parameter $p$. At this moment this can be considered as a coincidence or as an eventual new paradigm\,\footnote{See also references\cite{Duval}, \cite{Acedo}, \cite{Bootello}}.

Noteworthy is the fact that because the Hamiltonian is time dependent this model may have something to say about the anomalies of flybys. And last but not least it is also worthy to say that after so many theoretical physics models dedicated to explain the Pioneer's anomaly, now unnecessary \cite{Slava}, this one predicts a negligible contribution to this effect.

\vspace{1cm}
{\it Appendix}

\vspace{1cm}
Let us consider the following spherically symmetric space-time
model whose line-element is:

\begin{equation}
\label{1.12.2}
ds^2=-A^2dt^2+A^{-2}d\bar s^2, \quad d\bar s^2=M^2 dr^2+N^2r^2d\Omega^2
\end{equation}
where to start with we assume that:

\begin{equation}
\label{1.13}
A^{-1}=\left(\frac{r-m}{r+m}\right)^{-1/2}+(1+2p\,t)^{1/2}-1, \ \ G=c=1
\end{equation}

\begin{equation}
\label{1.14}
M=\frac{1}{\sqrt{1+p^2r^2}},\ \ \ N=\sqrt{1-\frac{m^2}{r^2}}
\end{equation}
If $p=0$ then (\ref{1.12.2}) is the Schwarzschild model, $r$ being the radial  Fock coordinate (x,y,z:harmonic):

\begin{equation}
\label{Sch}
A^2=\frac{r-m}{r+m}, \ \ M^2=1, \ \ N^2=1-\frac{m^2}{r^2}
\end{equation}
If $m=0$ then (\ref{1.12.2}) is Milne's flat space-time model:

\begin{equation}
\label{Milne}
A^2=1+2p\,t, \ \ M^2=\frac{1}{1+p^2r^2}, \ \ N^2=1
\end{equation}
But $t$ is not the global proper time that the model allows to use. It is the time that in both cases leads to a space model geometry:

\begin{equation}
\label{dsb2}
d\bar s^2=\frac{dr^2}{1+p^2r^2}+r^2d\Omega^2
\end{equation}
that is time independent and has constant curvature, thus fulfilling Helmholtz's free motion postulate.

The line-element (\ref{1.12.2}) can be considered in general as an approximate vacuum solution of Einstein's equations where the quality of the approximation depends on the relevant domains of $r$ and $t$ and the values of $m$ and $p$.

On the other hand, formal linear developments with respect to both $p$ and $m$, of (\ref{1.13}) and (\ref{1.14}) yield:

\begin{equation}
\label{Linear}
A^2=1-\frac{2m}{r}-2\left(1-3\frac{m}{r}\right)p\,t, \ \ M=N=1
\end{equation}
so that the central force per unit mass is:

\begin{equation}
\label{force}
f=-\frac{d\ln A}{dr}=-\frac{m}{r^2}(1-p\,t)
\end{equation}
At this formal approximation one has:

\begin{equation}
\label{Eins}
R^\alpha_\beta-\frac12\delta^\alpha_\beta=0
\end{equation}
and the non zero strict components of the Riemann tensor are:

\begin{eqnarray}
\label{Riem}
R^1_{.212}=-\frac{m}{r}(1+p\,t), \ \ R^1_{.313}=R^1_{.212}\sin^2\theta \\
R^4_{.242}=-\frac{m}{r}(1-3p\,t), \ \ R^4_{.343}=R^4_{.343}\sin^2\theta \\
R^1_{.414}=-\frac{2m}{r^3}(1-3p\,t), \ \ R^3_{.232}=2\frac{m}{r}(1+p\,t)
\end{eqnarray}

{\it Acknowledgements}

\vspace{1cm}

I wish to thank L. Acedo whose apt questions and comments helped me to write a better manuscript.

\end{document}